%% file: wilson.tex
\documentclass[12pt]{article}
\usepackage{epsfig}

\makeatletter
\@addtoreset{equation}{section}

\def\baselinestretch{1.2}
\def\href#1#2{#2}

\newcommand{\nl}{\hspace{-.65cm}}
\newcommand{\be}{\begin{equation}}
\newcommand{\ee}{\end{equation}}
\newcommand{\beq}{\begin{eqnarray}}
\newcommand{\eeq}{\end{eqnarray}}
\newcommand{\s}{*}
\newcommand{\z}{\zeta}
\newcommand{\inn}{\! \cdot \!}
\newcommand{\si}{\sigma}

\newcommand{\tr}{{\rm Tr}\,}
\newcommand{\bl}{\nl $\bullet$\ }

\begin{document}
\begin{titlepage}

\begin{flushright}
NSF-ITP-00-94\\
hep-th/0008075\\
\end{flushright}
\vfil\vfil

\begin{center}

{\Large {\bf  Observables of Non-Commutative Gauge Theories}}

\vfil

\vspace{5mm}

David J. Gross$^a$, Akikazu Hashimoto$^a$,  and N. Itzhaki$^b$\\

\vspace{10mm}

$^a$Institute for Theoretical Physics\\ University of California,
Santa Barbara, CA 93106\\
gross,  aki@itp.ucsb.edu\\

\vspace{10mm}

$^b$Department of Physics\\ University of California,
Santa Barbara, CA 93106\\
sunny@physics.ucsb.edu\\

\vfil
\end{center}

\begin{abstract}
We construct gauge invariant operators in non-commutative gauge
theories which in the IR reduce to the usual operators of ordinary
field theories (e.g.~$\tr F^2$). We show that in the deep UV the
two-point functions of these operators admit a universal exponential
behavior which fits neatly with the dual supergravity results.  We
also consider the ratio between $n$-point functions and two-point
functions to find exponential suppression in the UV which we compare
to the high energy fixed angle scattering of string theory.
\end{abstract}

\end{titlepage}
\renewcommand{\baselinestretch}{1.05}  

\section{Introduction}

In the study of gauge field theories, one is often interested in
computing the correlation functions of gauge invariant local operators
such as $ \tr F^2(x)$. These correlation functions contain a great
deal of information about the dynamics of the theory.  In particular,
they are localized probes of the gauge theory dynamics.  Localized
probes are especially interesting in the context of gauge theories on
non-commutative geometries. Non-commutativity introduces a new
physical scale to the problem, which gives rise to new physics which
one would like to study.

With respect to gauge invariant local operators non-commutative gauge
theory is quite different from its commutative counterparts. In a
non-commutative gauge theory, operators such as $\tr F^2(x)$, are not
gauge invariant, but rather they transform non-trivially. In order to
render such an operator gauge invariant out of these operators, one
must integrate over all space. In other words, the same operator in
momentum space, $\tr F^2(k)$, is gauge invariant only if $k=0$.  But
then the operator is not useful as a localized probe.

At first sight, it appears that gauge invariant operators which carry
non-zero momentum simply do not exist in non-commutative gauge
theories and thus the set of observables of NC gauge theories is much
smaller than in ordinary gauge theories.  There are different ways to
see that this possibility is not satisfactory.  Perhaps the most
compelling is due to the existence of a supergravity dual to NCSYM
\cite{hi,mr}.  The supergravity dual automatically captures the gauge
invariant dynamics of NCSYM. The excitations of supergravity modes are
not restricted to the zero momentum sector, implying that there is a
momentum dependent gauge invariant observables in the
theory. Moreover, the fact that the supergravity solution in the near
horizon limit asymptotes to $AdS_5 \times S_5$ implies that the
operators corresponding to these supergravity modes should approach
ordinary commutative SYM operators in the small momentum limit. This
raises the question: what are the non-commutative generalizations of
gauge invariant operator such as $\tr F^2(k)$?

Ishibashi, Iso, Kawai, and Kitazawa have recently constructed a set of
gauge invariant operators in non-commutative gauge theories carrying
non-vanishing momentum \cite{iikk} (See also
\cite{Ambjorn:1999ts,Ambjorn:2000nb,amns}). These authors showed that
an {\em open} Wilson line with momentum $p_\mu$ is gauge invariant if
the distance between the end-points of the line is
\be l^\nu = p_\mu \theta^{\mu \nu}. \ee
Roughly speaking, the gauge dependence of the open Wilson line is
canceled by the gauge dependence due to the momentum.

Just like the Wilson loops in ordinary gauge theories, these operators
constitute an over-complete set of gauge invariant operators of
non-commutative gauge theories \cite{amns}. However, it is not clear
at first sight how these operators correspond to the excitations of
the dual supergravity modes, and how these operators reduce to the set
of ordinary gauge invariant local operators in the small momentum
limit.  The goal of this paper is to address these issues.  It turns
out that a simple generalization of the IIKK construction will give
rise to the desirable set of operators.  These are the non-commutative
generalizations of the gauge invariant local operators! We will
investigate the correlation functions of these operators in
perturbation theory and compare the result with supergravity.

The organization of this paper is as follows. We will begin in section
2 by reviewing the construction of IIKK Wilson lines. We will then
describe how IIKK construction can be modified to give rise to the set
of operators that are ``closest'' to the ordinary gauge invariant
local operators. In section 3, we investigate the two-point function
of these operators in perturbation theory. We find a universal
exponential behavior in the UV.  In section 4, we find similar
behavior using the supergravity dual description.  In section 5, we
calculate the ratio between the $n$-point function and the two point
function to find an exponential suppression which we compare in
section 6 to the well known behavior of the high energy fixed angle
scattering of string theory.

The relation between open Wilson lines and supergravity was also
studied recently, using a different approach, in \cite{dasrey}.

\section{Gauge invariant operators in NCYM}

In this section we construct the non-commutative generalization of
gauge invariant operators. The construction is based on the open
Wilson lines of IIKK which we review below.  Let us first set the
notation that we use throughout the paper.  For simplicity, and to
avoid problems with Wick rotation, we take $\theta^{23}$ to be the
only non-vanishing component of the non-commutativity parameter. We
will take the gauge group to be $U(N)$. The action for this theory is
\be\label{action} S=\frac{1}{4}\int d^4 x\tr (F_{\mu\nu}(x) \s
F_{\mu\nu}(x)), \ee
where
\be F_{\mu\nu}=\partial_{\mu} A_{\nu}-\partial_{\nu} A_{\mu} +i
g (A_{\mu}\s A_{\nu} - A_{\nu}\s A_{\mu}),  \ee
and $\s$ is the familiar star product
\be f(x) \s g(x) \equiv
e^{\frac{i}{2}\theta^{\mu\nu}{\partial \over \partial x^\mu}
{\partial \over \partial y^\nu}} f(x) g(y)|_{x=y}.
\ee
The action is invariant under the non-commutative gauge
transformation,
\be A_{\mu}(x) \rightarrow U(x) \s A_{\mu}(x) \s U(x)^{\dagger}
-\frac{i}{g} U(x) \s \partial_{\mu} U(x)^{\dagger}, \ee
where $U(x)$ is the non-commutative gauge parameter, with $U(x) \s
U(x)^{\dagger}= 1$.  Under this transformation law, $\tr F^2(x)$ is
not gauge invariant but transforms according to
\be \tr F^2(x) \rightarrow \tr U(x) * F^2(x) * U^{\dagger}(x). \ee
However, integrating over all space will give rise to a gauge
invariant operator, since integrals of $*$-products can be cyclically
permuted.

Similarly a Wilson line can be generalized to non-commutative gauge
theories
\be W(x,C)= P_{\s} \exp \left(i g \int_0^1 d \sigma {d \z ^{\mu} \over
d \sigma} A_{\mu}(x+\zeta(\sigma))\right)  , \ee
where $C$ is the curve which parameterized by $\z^{\mu}(\sigma)$ with
$0\leq \si \leq 1$, $\zeta(0) = 0$, and $\zeta(1)=l$.  $P_{\s}$ denotes
path ordering with respect to the star product
\be W(x,C)= \sum_{n=0}^{\infty} (ig)^n
\int_0^1 d\sigma_1 \int_{\sigma_1}^1 d \sigma_2 ...
\int_{\sigma_{n-1}}^1 \!\!\!\!\!\! d \sigma_n \
\z^{'}_{\mu_1}(\si_1) ... \z^{'}_{\mu_n}(\si_n)
A_{\mu_1}(x+\z(\si_1))\s ... \s A_{\mu_{n}}(x+\z(\si_n)).  \ee
Under the gauge transformation, the Wilson lines transform according
to
\be\label{2} W(x,C) \rightarrow U(x) \s W(x,C) \s U(x+l)^{\dagger}.\ee
Just as in the ordinary gauge theories, an open Wilson line by itself
is not gauge invariant.  However, unlike in ordinary gauge theories,
closing the line does not make the Wilson line gauge invariant.  On
the other hand, in non-commutative gauge theories, one can construct a
gauge invariant operator out of the open Wilson lines in the following
way.  Consider the operator
\be\label{4}  W(k,C) =\int d^4 x~ \tr W(x,C) \s
e^{i k x}, \ee
which is simply the Fourier transform of the open Wilson line.  The
integration is over the base point while keeping the path fixed.
Under gauge transformations, this operator maps to
\be  W(k,C) \rightarrow
\int d^4 x ~\tr U(x) \s W(x,C) \s U^{\dagger}(x+l) \s e^{i
k x}. \label{gaugetransform}\ee
The reason why it is useful to Fourier transform to momentum space is
that in non-commutative geometry, $e^{ikx}$ is a translation operator.
That is,
\be e^{i k x} * f(x) = f(x +k \theta ) * e^{i k x}.  \ee
Eq.~(\ref{gaugetransform}) can, therefore, be written as
\be W(k,C) \rightarrow
\int d^4 x~ \tr U(x) \s W(x,C) \s e^{i k x} \s
U^{\dagger}(x+l - k \theta), \ee
and hence $W(k,C)$ is gauge invariant if $C$ satisfies the condition
\be\label{3} l^{\nu}=k_\mu \theta^{\mu\nu}.  \ee
Notice that this fixes only the distance between the end points of the
Wilson line but does not put any additional constraint on the shape of
the line.  Notice further that in the commutative limit ($\theta
\rightarrow 0$ keeping $k$ fixed), we find that gauge invariance
requires loops to be closed, as expected.

The set of open Wilson lines satisfying the condition (\ref{3})
constitutes an over complete set of gauge invariant operators, just
like the closed loops in ordinary gauge theories \cite{amns}.  We will
now describe how one constructs a convenient set of gauge invariant
operators which is a natural generalization of the standard local
gauge theory operators in the commutative limit.

\begin{figure}
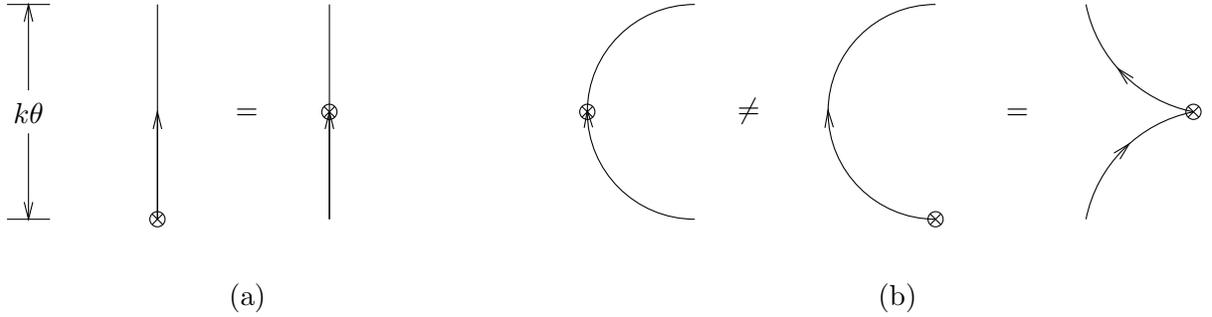

\centerline{\input operators.pstex_t}
\caption{Moving the location of the local operator on the Wilson line
does not change the definition of the gauge invariant operator only if
the line is straight (a). For a generic line one gets a different
gauge invariant operator (b).  \label{figaa}}
\end{figure}

Consider an operator which consists of the usual local operator
attached at one end of the Wilson line with non-vanishing momentum. An
example of such an operator is
\be
\tr \tilde F^2(k)=\int d^4 x ~\tr  F^2(x) \s W(x,C) \s e^{ikx}.
\ee
As long as $C$ satisfies the condition (\ref{3}), such an operator
will be gauge invariant. In fact, any local operator in the adjoint
representation can be attached at the end of the Wilson line
satisfying (\ref{3}) to give rise to a gauge invariant operator. See
figure \ref{figaa} for an illustration.  Clearly, the set of all
operators consisting of all local operators attached to Wilson lines
of all shapes is an over-complete set.  Since information about the
shape of the lines can be absorbed into the local operators, one can
drastically reduce the redundancies in the set of operators by
requiring the Wilson line to take on a definite shape, but allowing
arbitrary local operators to be attached at its endpoint.  The
simplest choice is to take these lines to be a straight line
satisfying the condition (\ref{3}).  Just like the closed straight
Wilson loop on a torus, such an operator acts as an order parameter
for the large gauge transformations.

It may seem somewhat ad-hoc to attach the operator at one end of the
Wilson line. Why not attach it in the middle, or smear it evenly
throughout the line?  One could have taken any of these prescriptions
in defining the set of operators.  It turns out, in the case of a
straight Wilson line, that all of these prescriptions give rise to the
same operator\footnote{We thank J. Maldacena for pointing this out to
us.}. This is yet another reason why the straight Wilson lines are the
most natural lines to consider.  This follows from the fact that
\be \int d^4 x ~ {\cal O}(x) * W(x,C) * e^{i k x} = \int d^4 x ~
W^\dagger (x,-C_2) * {\cal O}(x) * W(x,C_1) * e^{i k x} \ , \ee
where
\be   C = C_1 + C_2 \ .  \ee
Only for the straight line will the shape of the line remain invariant
with respect to this transformation.  See figure \ref{figaa} for an
illustration.

To summarize: For any local operator of ordinary gauge theories ${\cal
O}(x)$ in the adjoint of the gauge group we have constructed a
non-commutative generalization
\be\label{5}
\tilde{{\cal O}}(k) = \tr \int d^4 x\  {\cal O}(x) \s
P_{\s} \exp\left(i g \int_{C} d \z^{\mu} A_{\mu}(x+\z)\right) \s e^{i k x},
\ee
where $C$ is a straight path
\be
\z^{\nu}(\si)=  k_{\mu} \theta^{\mu\nu} \si,\qquad 0\leq\si<1.
\ee
The tilde is used to emphasize the fact that we have attached a Wilson
line to the operator. This should be thought of as the generalization
of Fourier transforms of gauge invariant operators to the
non-commutative case.

\nl Several comments are in order regarding the set of operators  (\ref{5}).

\bl Due to the antisymmetry of $\theta^{\mu \nu}$, $k_\mu$ and $l^\nu$
are orthogonal to one another. In other words, the Wilson line is
extended in the direction transverse to the momentum.

\bl For small values of $k$ or $\theta$, the length of the Wilson line
goes to zero and (\ref{5}) is reduced to operators in the ordinary
field theory
\be\label{6}
\tilde{{\cal O}}(k) \rightarrow {\cal O}(k) = \int d^4 x\, {\cal O}(x)
e^{i k x}. \ee

\bl The fact that $\tilde {\cal O}(k)$ contains Wilson lines whose
length depends on $k$ means that one should not think of these
operators as the ``same'' operator with different momentum as we
usually do in commutative theories. One should think of $\tilde {\cal
O}(k)$ as genuinely different operators at different momentum. One
should therefore not expect to obtain a local operator by Fourier
transforming to position space. This is closely related to the fact
that these operators require momentum dependent regularization in
perturbation theory and in the supergravity dual \cite{mr}. They are
simply different operators.

\bl At large values of $k$, the length of the Wilson line becomes
large as dictated by the non-commutativity relation.  In this limit,
the operator is dominated by the Wilson line regardless of what
operator is attached at the end. We therefore expect the correlation
function of these operators to exhibit a universal large $k$ behavior.

In section 3 we will explain the computation of correlation functions
for these operators in perturbation theory and elaborate on their
universal features at large momentum.

\subsection{Operator formulation}

For completeness we describe in this sub-section an alternative
derivation of such gauge invariant observables using the operator
formulation of non-commutative gauge theory.  We introduce
\be c = {1\over{\sqrt{2\theta}}} \left( x^2- i x^3
\right),\quad c^{\dagger} = {1\over{\sqrt{2\theta}}} \left( x^2 + i
x^3 \right),\ee
which obey
\be [ c , c^{\dagger} ] = 1. \ee
Since $c, c^{\dagger}$ satisfy the commutation relations of the
annihilation and creation operators we can identify functions
$f(x^2,x^3)$ with operator functions of $\hat c$, $\hat{c}^{\dagger}$
acting in the standard Fock space of the creation and annihilation
operators by Weyl ordering, as defined by
\beq && f(x)=f\left( z=x^2- ix^3, \bar z=x^2+ix^3\right)\\
\nonumber && \mapsto {\hat f}({\hat c}, {\hat c}^\dagger) = \int
\frac{d^2 x d^2 p}{(2\pi)^2} f(x) \exp \left( i {\bar p}(
\sqrt{2\theta}\, {\hat c} - z ) +i p ( \sqrt{2\theta}\, {\hat c}^\dagger
- {\bar z}) \right) .  \eeq
It is easy to see that if $f \mapsto {\hat f},$ and $g \mapsto {\hat
g}$, then $f * g \mapsto {\hat f}{\hat g} \ ,$ and
\be \int dx^2 dx^3 f(x) = 2 \theta\, {\rm tr}[{\hat f}({\hat c}, {\hat
c}^\dagger)],\ee
where ${\rm tr}$ is a trace over the Fock space. Translations in the
Hilbert space are generated by ${\hat\partial_i}$, where
\be\label{19} 2\theta\,\hat\partial_2= \sqrt{2\theta}\, ({\hat
c}-{\hat c}^\dagger)=-2i {\hat x}^3, \quad 2\theta\, \hat\partial_3=i
\sqrt{2\theta}\,({\hat c}+{\hat c}^\dagger)=2i {\hat x}^2 .  \ee
Thus, if $ f(x) \mapsto {\hat f}$, then $f(x+l) \mapsto \exp(l \cdot
\hat \partial ){\hat f}\exp(-l \cdot \hat\partial)$ .  The covariant
derivative of a $U(N)$ gauge field is then represented as the
operator,
\be
{\hat D}_2 = -i\hat\partial_2 +g{\hat A}_2, \qquad
{\hat D}_3 =- i\hat\partial_3 +g{\hat A}_3 , \ee
where ${\hat A}_\mu$ are $N\times N$ Hermitian matrix operators in the
Fock space that represent the components of the gauge field.  Under a
gauge transformation the covariant derivative transforms like a field
in the adjoint representation:
\be{\hat D}_\mu \to{\hat U }{\hat D}_\mu{\hat U}^\dagger \quad{\hat
U}^\dagger{\hat U }={\hat U }{\hat U }^\dagger= 1\ . \ee

In this formalism it is easy to see why observables localized in
position space are not gauge invariant.  This is because translations
of operators in the non-commutative directions are equivalent, up to
constant shifts of the gauge field, to gauge transformations.
Translations are generated by the operators $\hat\partial$, defined in
eq.~(\ref{19}). Thus the translation, by an amount $l=(l^2,l^3)$, of a
field, ${\hat \Phi}$, in the adjoint representation and of ${\hat
A}_\mu$, are given by
\be \hat\Phi \to \exp(l\cdot \hat\partial){\hat \Phi} \exp(-l\cdot
\hat\partial), \quad {\hat A}_\mu \to \exp(l\cdot \hat\partial){\hat
A}_\mu \exp(-l\cdot \hat\partial) .  \ee
This is a gauge transformation of the Higgs field, ${\hat \Phi} \to
{\hat U}(l) {\hat \Phi} {\hat U}^\dagger(l)$, where
\be
{\hat U}(l)= \exp(l\cdot \hat \partial)= \exp(i l^i
\theta^{-1}_{ij}{\hat x}^j ) .
\ee
Acting on the gauge field this transformation yields
\be {\hat A} \to {\hat U}(l) {\hat A} {\hat U}^\dagger(l) = \left[ {\hat
U}(l) ({\hat A}-{\hat c}^\dagger) {\hat U}^\dagger(l) +{\hat
c}^\dagger\right] + {\hat U}(l)[ {\hat c}^\dagger, {\hat
U}^\dagger(l)]\equiv \delta_1 {\hat A} +\delta_2 {\hat A} \ . \ee
The first term, $\delta_1 {\hat A}$, is a gauge transformation and the
second term, $\delta_2 {\hat A}$, is a constant shift of the gauge
field,
\be \delta_2 {\hat A} = {\hat U}(l)[ {\hat c}^\dagger, {\hat
U}^\dagger(l)] =-(l^2+i l^3)\ . \ee
Both of these, gauge transformations and constant shifts of the gauge
field, are symmetries of the action.

What is unusual about non-commutative gauge theories is that {\it
translations in the non-commutative directions are equivalent to a
combination of a gauge transformation and a constant shift of the
gauge field.} This explains why in NC gauge theories there do not
exist local gauge invariant observables in position space, since by a
gauge transformation we can effect a spatial translation! This is
analogous to the situation in general relativity, where translations
are also equivalent to gauge transformations (general coordinate
transformations) and one cannot construct local gauge invariant
observables. The fact that spatial translations are equivalent to
gauge transformations (up to global symmetry transformations) is one
of the most interesting features of NC gauge theories. These theories
are thus toy models of general relativity---the only other theory that
shares this property.  However, unlike the case of general relativity,
it is easy enough to derive a large set on non-local, gauge invariant,
observables.

Suppressing the dependence on the commuting coordinates, a complete
set of gauge invariant loops is given by
\be
\tr{\rm tr}\left[ \prod_i \exp\left(i {\hat D}\cdot l_i\right) \right].
\ee
The double trace, ($\tr$) over the $N\times N$ matrices and (${\rm
tr}$) over the Fock space states, is necessary to ensure gauge
invariance. A Wilson line segment, from $x$ to $x+l$, is represented
by the operator
\beq && W(x, x+l)= P_* \exp \left(i g \int_x^{x+l} d \z ^{\mu}
A_{\mu}( \zeta)\right) \mapsto \cr
&& {\hat W}({\hat x}, {\hat x}+l) \equiv \exp({\hat x \cdot {\hat
\partial}}) \exp\left(i {\hat D}\cdot l \right)\exp(-({\hat x}+l) \cdot
\hat \partial). \eeq
Therefore, using eq.~(\ref{19}) we get,
\beq &&\tr{\rm tr}\left[ \prod_i \exp\left(i {\hat D}\cdot l_i\right)
\right] =\tr{\rm tr}\left[ {\hat W}(0,l_1){\hat W}(l_1,l_1+l_2) \dots
{\hat W}(L-l_{n},L) \exp i {L\cdot \hat \partial} \right] \nonumber \\ 
&& \mapsto \int d x^2 dx^3 \, \tr P_*\exp \left(i g \int_{C(l_i)} d \z
^{\mu} A_{\mu}( \zeta ) \right) \exp\left(
iL^i\theta^{-1}_{ij}x^j\right)\ , \eeq
where $L = \sum_i l_i$ and ${C(l_i)}$ is the contour from $x$ to $x+L$
composed out of the $n$ line-segments $l_i$. These are precisely the
operators considered previously.

\section{Correlation function of gauge invariant observables}

Let us take the coupling constant to be small so that we can use
perturbation theory.  In the next section we compare our results with
the supergravity results which are valid at large coupling to find a
nice agreement. We consider ${\cal N}=4$ $U(N)$ theory which contains
adjoint scalars and fermions in addition to the gauge bosons. We will
take the 't Hooft limit so that only the planer graphs contribute even
though this is not crucial for the main results of this section.

\begin{figure}
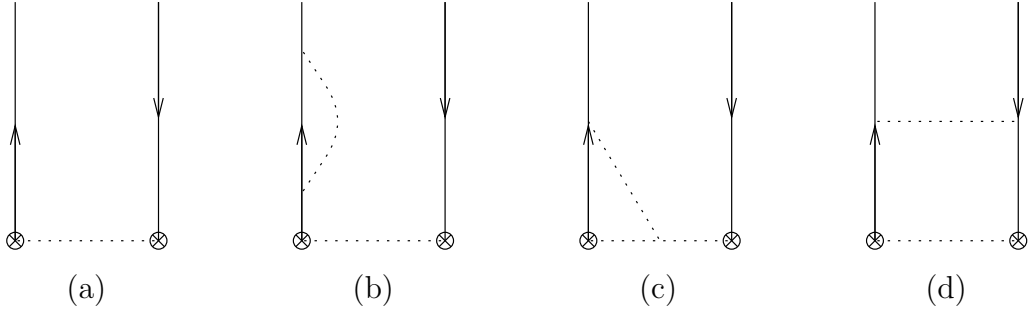

\centerline{\input feynman.pstex_t}
\caption{Leading contributions to the two-point functions.  (a) is the
$g^0$ order which agrees with the ordinary gauge theory results.  (b),
(c) and (d) are the non-commutative corrections at order $g^2$.
\label{figa}}
\end{figure}

Effects due to non-commutativity are weak in the infra red and grows
as one increases the energy. In the extreme ultra violet, the
correlation functions exhibit a universal exponential dependence on
the energies which does not depend on the choice of ${\cal O}(x)$. Let
us illustrate this with a concrete example. Perhaps the simplest
operator to consider is to take ${\cal O}(x) = \tr \phi(x)$. The
relevant corresponding non-commutative operator is
\be\label{a1}
\tr \tilde \phi(k) = \tr \int d^4 x\,  \phi(x)
\s P_{\s} \exp\left(i g \int_0^1 d \sigma {\z'}^{\mu}
A_{\mu}(x+\z(\sigma))\right) \s e^{i k x},
\ee
with
\be
\z^{\nu}(\si)=k_\mu \theta^{\mu\nu} \si.
\ee
In perturbation theory, one can expand (\ref{a1}) order by order in
$g$,
\beq
\tr \tilde \phi(k) &=&
\tr \left( \phi(k) + ig  \int d^4 x \int  d \z_1 \,   \phi(x) * A(x + 
\z_1) * e^{i k x} \right.\nonumber \\
&& \left.+ (ig)^2 \int d^4 x \int d \z_1 d \z_2  \,\phi(x) * A(x + 
\z_1)* A(x + \z_2)) * e^{i k x} + \ldots \right).
\eeq
Under complex conjugation, the path ordering is reversed
\beq
\tr \tilde \phi^\dagger(k)
&=& \tr \left( \phi(k) - ig  \int d^4 x \int  d \z_1 \,   A(x + \z_1) 
*  \phi(x) * e^{i k x} \right.\nonumber \\
&& \left.+ (ig)^2 \int d^4 x \int d \z_1 d \z_2  \,  A(x + \z_2)) * 
A(x + \z_1)*\phi(x) * e^{i k x} + \ldots \right).
\eeq
To facilitate the perturbative calculation using standard Feynman
rules, it will be convenient to express the fields in momentum space
\be \phi(x) = \int {d^4 p_0 \over (2 \pi)^4} \phi(k) e^{-i p_0 x}, 
\qquad A_\mu (x+\zeta_i) = \int {d^4 p_i \over (2 \pi)^4} A_\mu(p_i) 
e^{-i p_i (x + \zeta_i)}. \ee
The $*$-product will only act on the exponential factor and generate
the usual non-commutative phase factors. Integrating over $x$ will
constrain $\sum p_i =k$, and gives\\
\parbox{\hsize}{
\beq
\tr \tilde \phi(k) &=&
\tr \left( \phi(k) + ig  \int {d^4 p_1 \over (2\pi)^4} \int  d \z_1 
\,   \phi(k - p_1 )  A(p_1) e^{i k \theta p_1 /2}\right.\nonumber \\
&&+ (ig)^2  \int {d^4 p_1 \over (2\pi)^4}  \int {d^4 p_2 \over 
(2\pi)^4} \int d \z_1 d \z_2  \,\phi(k-p_1-p_2) A(p_1) A(p_2)  e^{i 
(k \theta p_1 + k \theta p_2 + p_1 \theta p_2)/2} \nonumber \\
&& \left.\rule{0ex}{3.5ex} + \ldots \right) . \label{expansion}
\eeq
}
An interesting dynamical quantity to study is the two point function
\be
\langle \tr \tilde{\phi}(k_1) \tr \tilde{\phi}^\dagger(k_2) \rangle.
\ee
The leading order contribution at order $g^0$ comes from the diagram
(a) in figure \ref{figa}. This is simply the commutative result
\be \langle \tr \phi(k_1) \tr \phi(-k_2) \rangle = {N \over k_1^2} \,
(2 \pi)^4 \delta^4 (k_1-k_2). \ee
We will drop the momentum conserving $\delta$-function from most of
the discussion below.

At order $g^2$, one gets a contribution from diagrams (b), (c), and
(d) in figure \ref{figa}. Let us evaluate these diagrams explicitly.
The contribution to diagram (b) comes from crossing the first and the
third term of (\ref{expansion}). The gauge bosons in the third term of
(\ref{expansion}) is contracted to each other, so that we can replace
\be A(p_1) A(p_2) \rightarrow
(2 \pi)^4 \delta^4 (p_1 + p_2) {1 \over p_1^2}. \ee
This will cause the non-commutative phase factors to cancel, and the
remaining $p$ integrals can be done explicitly to give
\be {(ig)^2 \over k^2} \int d \z_1 d \z_2 {1 \over 4 \pi^2 (\z_1 -
\z_2)^2}. \ee
The integral diverges when $\z_1$ and $\z_2$ approach one another. If
we regulate this integral at ultra-violet scale $\Lambda$, we find
\be {\rm (b)} = { g^2 N^2 \over 4 \pi^2 k^2} |k \theta| \Lambda, \ee
as the contribution from the diagram (b).

Similar techniques can be used to compute the contribution from the
diagram (c). The phase factor in (\ref{expansion}) cancels the phase
factor from the three point vertices, and we find
\be {\rm (c)} = {(ig)^2 N^2 \over k^2} \int {d^4 p_1 \over (2 \pi)^4}
{1 \over (k-p_1)^2} {1 \over p_1^2} \int_0^1 d \sigma_1\, i ( k \theta
p_1 ) e^{- i( k \theta p_1) \sigma_1}. \ee
After doing the $p_1$ and $\sigma_1$ integrals, we find that (c)
scales according to
\be {\rm (c)} \sim
\left\{ \begin{array}{lcl}
\parbox{1.5in}{$${ g^2 N^2 \over k^2} \log( |k \theta| |k|)$$},
&\qquad& \mbox{IR}\\
\parbox{1.5in}{$${ g^2 N^2 \over k^2} $$}, &\qquad & \mbox{UV}
\end{array} \right. \label{diagramc}\ee
where the IR and UV refers to $|k||\theta k| \ll 1$ and $|k||\theta k|
\gg 1$, respectively.  Finally, from diagram (d), we find the
contribution
\be {\rm (d)} = - (ig)^2 N^2 |k \theta|^2 \int {d^4 p_1 \over (2
\pi)^4} {1 \over (k-p_1)^2} {1 \over p_1^2} \int_0^1 d\sigma_1
\int_0^1 d \sigma_2 \, e^{-i (k \theta p_1) (\sigma_1 - \sigma_2)}.
\ee
Doing the $p_1$ and $\sigma_{1,2}$ integral shows that (d) scales as
follows
\be {\rm (d)} \sim
\left\{ \begin{array}{lcl}
\parbox{2in}{$$ {g^2 N^2 \over k^2 } (|k| |k \theta|)^2 \log(|k| |k
\theta|) $$}, & \qquad&
\mbox{IR}\\
\parbox{2in}{$${g^2 N^2 \over k^2} |k| |k \theta| $$}, & \qquad &
\mbox{UV.}
\end{array} \right. \label{diagramd} \ee
We see that in the IR the corrections are small, as expected.  The
more interesting question is how these correlation functions behave at
energies much larger than the non-commutativity scale.  From the
behavior of (\ref{diagramc}) and (\ref{diagramd}) at large momentum,
we see that the leading contribution is due to diagram (d). The
perturbative correction due to (\ref{diagramd}) is controlled by the
dimensionless parameter
\be g^2 N |k| |\theta k|. \ee
Even for small coupling constant, at sufficiently large $k$, this
parameter will be greater than one. Therefore, contributions from
higher order diagrams must be taken into account. Fortunately, it is
possible to identify the diagrams which dominate at each order in the
perturbative expansion and perform the resummation to reliably compute
the behavior for large $|k | |\theta k|$.

To gain some intuition on why the resummation is possible it is useful
to recall that something very similar happens in the calculation of
quark anti-quark force using Wilson loops in ordinary commutative
gauge theories.  There, one computes the expectation value of a
rectangular loop of size $T \times L$ where $T$ is in the time
direction and $L$ is the distance between quarks and $T\gg L$.  If the
gauge theory is in the Coulomb phase, the Wilson loop expectation
value takes the form
\be W(T,L) = \exp(g^2 N T / L) = 1 + {g^2 N T \over L} + {1 \over 2! }
\left( {g^2 N T \over L} \right)^2 +{1 \over 3! } \left( {g^2 N T
\over L} \right)^3 + \ldots \label{series}.\ee
Again it seems that even for small 't Hooft coupling, the expansion
parameter becomes too big for large enough $T$.  But in fact, the
exponential form of (\ref{series}) is obtained using perturbation
theory by summing over ladder diagrams which dominate when $T\gg L$.

The problem of computing the two point function of the open Wilson
lines at large momentum is very similar to the problem of computing
the expectation value of rectangular Wilson loops. The length of the
line $| k \theta |$ plays the role of $T$ and the momentum $|k|$ plays
the role of $1/L$. It is clear that taking $T \gg L$ corresponds to
$|k| |\theta k| \gg 1$.  In the calculation of $\langle \tilde
\phi(k_1) \tilde \phi^\dagger(k_2) \rangle$ to order $g^2$, we saw
that the leading large momentum contribution came from the ladder
diagram, and we expect this pattern to persist to higher orders. This
suggests that we should resum the ladder diagrams in order to evaluate
the large momentum behavior of these expectation values, just as one
does in the rectangular Wilson loops.

\begin{figure}
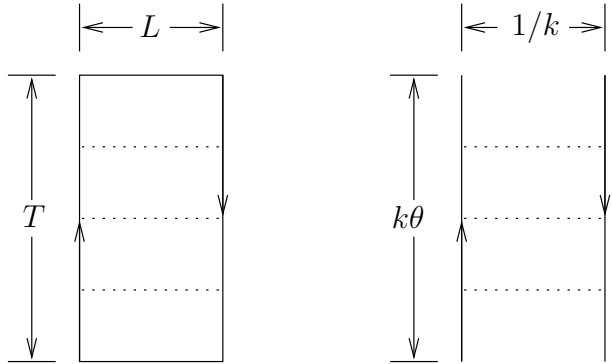

\centerline{\input loop.pstex_t}
\caption{Ladder diagram contributions to the rectangular Wilson loop
in ordinary field theories and to the open Wilson lines in
non-commutative gauge theories. \label{figb}}
\end{figure}

In fact, one can compute explicitly the leading large $k$ contribution
from the planer ladder diagrams to the two point correlation function
of the non-commutative gauge invariant operators. The general features
will not depend very much on the choice of operator ${\cal O}(x)$ one
attaches to the Wilson line. For the sake of simplicity, let us take
${\cal O}(x)$ to be an identity operator so that our non-commutative
gauge invariant operator is a pure Wilson line. We then find that
$\langle W(k) W^{\dagger}(k) \rangle$ receives contribution from the
$n$-th ladder diagram of the form
\be g^{2n} N^n  \int d^3 r {2 \over (n+1)!} \left({1 \over 4 \pi
|r|}\right)^n |k \theta|^{n+1} e^{i k r} \label{ladders} .\ee
The steps leading to this expression are summarized in the appendix
A. The integrand of (\ref{ladders}) will resum to an exponential. The
integral over $r$ is essentially a Fourier transform.  However, there
are some subtleties associated with this Fourier transform which we
will now explain\footnote{These subtleties are not special to
non-commutative theories and also appear if one tries to Fourier
transform eq.~(\ref{series}) in ordinary gauge theories}.  For $n \le
1$, the integral over $r$ converges fine, but for $n \ge 2$, the
integral receives strong divergent contributions from the region near
$r=0$. One can regulate this integral by introducing a small distance
cut-off $r > 1/\Lambda$.

In momentum space, we are only interested in terms which are
non-analytic in $k^2$.  Terms analytic in $k^2$ corresponds to contact
terms in the position space and do not contribute when the operators
are separated, say, in one of the commutative directions. In fact, we
could have chosen to Fourier transform only in the $x_2$ and $ x_3$
directions and continue to work with position space in the $x_0$ and
$x_1$ directions. It turns out that all terms which diverge as we
remove the cut-off $\Lambda$ are analytic in $k^2$. We are therefore
justified in dropping these terms. The non-analytic contribution to
the Fourier transform of $x^{-n}$ is simply
\beq
n  \ge 2\ ~~ {\rm even}  && -{2 \pi^2 i \over   (-k^2)^{3/2}}  {(-k^2)^{n/2}
  \over (n-2)!} \label{fouriera},\\ \nonumber
n   \ge 3\ ~~ {\rm odd}&&  -{2 \pi \over  (-k^2)^{-3/2}}
  \log (k^2/\Lambda^2) {(-k^2)^{n/2} \over (n-2)!}. \label{fourierb}
\eeq
Applying (\ref{fouriera}) and (\ref{fourierb}) to (\ref{ladders}), we
find that the non-analytic contribution to the ladder diagrams are
given by
\begin{eqnarray}
n \ge 2 \ ~~ {\rm even}  && -{4 \pi^2 i  |k \theta|  \over
  (-k^2)^{3/2}} {z^n  \over (n+1)! (n-2)!},
\\ \nonumber
n   \ge 3 ~~ {\rm odd}  && -{4 \pi |k \theta| \over   (-k^2)^{3/2}}
  \log (k^2/\Lambda^2) {z^n   \over (n+1)! (n-2)!},
\end{eqnarray}
where
\be
  z = \left( g^2 N |k \theta| \sqrt{-k^2} \over 4 \pi \right).
\ee
This  can be resummed to
\beq
n \ge 2 \ {\rm even}  && -{2 \pi^2 i  |k \theta| \over (-k^2)^{3/2}}
\left( \sqrt{z} I_3(2 \sqrt{z}) +  \sqrt{z} J_3(
2 \sqrt{z})\right ) \label{even}, \\ \nonumber
n \ge 3\  {\rm odd}  &&
-{2 \pi   |k \theta| \over  (-k^2)^{3/2}} \log (k^2/\Lambda^2) 
\left( \sqrt{z}I_3(2 \sqrt{z}) - \sqrt{z}
J_3(2 \sqrt{z}) \right) \label{odd}.
\eeq
Particularly simple quantity one can consider is the imaginary part of
this expression when $k^2 <0$. They are
\beq
n \ge 2 \ {\rm even}  && -{2 \pi^2   |k \theta | \over  (-k^2)^{3/2}}
\left( \sqrt{z} I_3(2 \sqrt{z}) + \sqrt{z} J_3(2 \sqrt{z})\right ),
\\ \nonumber
n \ge 3\ {\rm odd} && -{2 \pi^2 |k\theta | \over (-k^2)^{3/2}} \left(
\sqrt{z} I_3(2 \sqrt{z}) - \sqrt{z} J_3(2 \sqrt{z}) \right).  \eeq
Combining the odd and the even parts give rise to a very simple
expression
\beq
{\rm Im} \langle W(k) W^{\dagger} (k) \rangle = -{4 \pi^2
|k \theta| \over (-k^2)^{3/2}}\sqrt{z}  I_3(2 \sqrt{z}).
\eeq
Using the fact that $ I_\nu(x) \approx {e^x / \sqrt{2 \pi x}} $ we
find that this expression has an asymptotic behavior
\be {\rm Im} \langle W(k) W^{\dagger} (k) \rangle \approx {2
\sqrt{\pi}^{3/2} |k \theta| \over (-k^2)^{3/2}} z^{1/4} e^{2
\sqrt{z}}. \ee
What we found is that the two point function of gauge invariant
operators in non-commutative gauge theories {\em grows} exponentially.
So far we have ignored the contribution from the UV divergent diagram
(a) in figure \ref{figa}. In the case of the square Wilson loop, these
diagrams resum like $e^{- g^2 N \Lambda T / 4 \pi^2}$. The same is
happening here and hence the two point function of the non-commutative
Wilson lines behaves like
\be \langle W(k) W^\dagger (k) \rangle \sim \exp \left( - {g^2 N |k
\theta| \Lambda \over 4 \pi^2} + \sqrt{g^2 N |k \theta| |k| \over 4
\pi }\right),
\label{universal}\ee
at large momenta. The main contribution to this behavior comes from
the long Wilson line which is part of all the gauge invariant
operators carrying large momenta.  Therefore, this is a universal
property of gauge invariant operators in non-commutative gauge
theories.  Attaching an operator to the Wilson line will only modify
the power of momentum in front of the universal multiplicative
exponential factor.

There are several corrections to this leading result which one should
take into account. One is the finite $k$ correction which is
suppressed by $1/|k| |\theta k|$.  Such a correction is the analog of
the $L/T$ corrections to the rectangular Wilson loops in ordinary
gauge theories and corrects the ladder approximation.  The other
corrections are due to finite 't Hooft coupling.  These corrections
come about from summing up ladder diagrams which are sub-leading in
the coupling constant.  The effect of these corrections is to replace
the 't Hooft coupling in the exponent by some function of the coupling
constant while leaving the dependence on $|k \theta | $ and $|k|$ (or
$T$ and $L$ in the ordinary field theory case) intact.  At large
coupling it was shown in \cite{rey,malda}, in the ordinary field
theory case, that this function is $\sqrt{g^2 N}$.  In the next
section we shall find a similar behavior in our case.

\section{Supergravity dual of NCSYM}

So far we have concentrated on the properties of the non-commutative
gauge invariant operators from the point of view of field theory.  We
found a remarkable universal behavior for the two point functions at
large momentum due to the presence of the large Wilson line in these
operators.  We will show in this section that much of these features
can also be seen from the dual supergravity description of NCSYM.  The
supergravity dual of NCSYM was found in \cite{hi,mr} and takes the
form
\beq\label{lk}
ds^2 &=& \alpha' \left\{
\frac{U^2}{\sqrt{\lambda}}(-dt^2+dx_1^2) +\frac{\sqrt{\lambda}
U^2}{\lambda + U^4 \Delta^4}(dx_2^2+dx_3^2)
+\frac{\sqrt{\lambda}}{U^2}dU^2 +\sqrt{\lambda}d \Omega_5^2 \right\},
\\ \nonumber
e^{\phi}&=&\frac{\lambda}{4\pi N}
\sqrt{{\lambda \over \lambda + \Delta^4 U^4}}.
\eeq
We have only written the background metric in string frame and the
dilaton here. $\lambda = 2 \pi g^2 N$ is the 't Hooft coupling
constant. We have taken the non-commutativity parameter to be
non-vanishing only in the $\theta^{23} = 2 \pi \Delta^2$ component.

In the AdS/CFT correspondence, one compares the fluctuations of the
supergravity background on the supergravity side to the correlation
function of gauge invariant operators on the field theory side. In
order to make such a comparison precise, one must understand the
fluctuations on the supergravity side along the lines of
\cite{Kim:1985ez}.  Unfortunately, the presence of a non-trivial
dilaton and other supergravity field background gives rise to more
mixing of the small fluctuations than in the $AdS_5 \times S_5$ case.
This mixing has to be diagonalized.  The complete treatment of this
issue appears to be a rather non-trivial task. Note however that in
the small $U$ limit, the supergravity background approaches $AdS_5
\times S_5$, and all the effect of the mixing will be suppressed by
$\lambda^{-1} \Delta^4 U^4 \sim \lambda \Delta^4 E^4$ where we have
used the relation $E = U / \sqrt{\lambda}$.  The mixing of the
supergravity fluctuations and the mixing of the gauge invariant
operators due to the Wilson line attached to them are related.  Though
hard to handle, this mixing is an important ingredient in the duality
between NCSYM and string theory on the relevant background (\ref{lk}).

\subsection{Non-decoupling of the $U(1)$ }

One effect of the mixing is that the $U(1)$ sector of $U(N) = U(1)
\times SU(N)/Z_N$ no longer decouples from the $SU(N)/Z_N$ when the
effect of non-commutativity is taken into account. Let us explain how
to see this explicitly in the language of the supergravity dual.
First recall that $AdS_5 \times S_5$ is dual to $SU(N)/Z_N$ rather
than $U(N)$ gauge theory. There are many different ways to argue this
\cite{Witten,og,ofer,Witten2}.  Perhaps the most intuitive argument is
the fact that gravity couples to everything, so the $U(1)$ cannot be
part of the dynamics of the supergravity dual \cite{Witten}. A
concrete way to see this is to consider $U(N)$ SYM on $R^2 \times T^2$
with non-zero 't Hooft flux along the $T^2$.  All of the energies
associated with the 't Hooft flux is contained in the $U(1)$ part of
the gauge group \cite{th}. From the point of view of supergravity,
turning on a 't Hooft flux corresponds to turning on a constant
$B_{NS}$ background \cite{Witten2}.  (See also \cite{hi2} for a
discussion of 't Hooft fluxes in supergravity in relation to Morita
equivalence.) The fact that a constant $B_{NS}$ can be turned on
without any cost in energy implies that the supergravity dual knows
only about the $SU(N)/Z_N$.

In the case of the supergravity dual of the NCSYM, on the other hand,
one can not turn on $B_{NS}$ without costing energy.  This is due to
the Chern-Simons term in type IIB supergravity
\be \int F_5\wedge F_3 \wedge B_{NS}, \ee
where $F_5$ and $F_3$ are the Ramond-Ramond five and three form field
strengths, respectively.  In pure $AdS_5\times S_5$ this term does not
prevent us from turning on a constant $B$ field since
$F_3=0$.\footnote{On the other hand this term is very important for
the understanding of many related issues in $AdS_5$ including the
existence of a baryon vertex operator \cite{og} and the quantization
of $B_{NS}$ and $B_{RR}$ \cite{ofer,Witten2}.} In the non-commutative
case, $F_3$ is non-zero, so a constant $B$ field will couple to the
background and cannot be turned on without costing energy.  In pure
$AdS_5 \times S_5$, constant $B_{NS}$ is part of the singleton
multiplet. So it decouples from the rest of the fields in the small
$U$ region where the background (\ref{lk}) is effectively $AdS_5
\times S_5$, but they mix at large $U$. This is the sense in which the
singleton fields and other supergravity modes are mixed, which can be
viewed as the mixing of the $U(1)$ from the field theory side.
 
\subsection{Two point function and the absorption cross section}

Clearly, disentangling the mixing of supergravity modes is a highly
non-trivial task.  However, since we are primarily interested in
understanding the properties of the correlation functions which are
universal, let us consider a generic fluctuation on the supergravity
side. In a given supergravity background, there are generally scalars
which couple minimally to the background.  The field equation for such
a field is given by
\be {1 \over \sqrt{g^E }}\partial_\mu \sqrt{g^E}  g_E^{\mu \nu}
\partial_\nu \Phi(x,U,\Omega) = 0 \ee
where $g^E$ is the Einstein metric $ g_{\mu \nu}^E = e^{-\phi/2}
g_{\mu \nu}^S $.  Working with a fixed momentum mode
\be \Phi(x,U,\Omega) = e^{i k x} \Phi(U), \ee
and writing in the form of a Schroedinger equation by redefinition of
the fields $\Phi(U) = U^{-5/2} \Psi(U)$, one finds
\be - {\partial^2 \over \partial U^2} \Psi(U) + V(U) \Psi(U) = 0 ,
\qquad V(U) = {\lambda k^2 \over U^4} + {15 \over 4 U^2} + \left|{k
\theta \over 2 \pi} \right|^2 .\ee
If $k^2 < 0$ the region near $U=0$ is classically allowed. Therefore,
one can tunnel through the barrier from $U=\infty$. The tunneling
amplitude will then correspond to the imaginary part of the two point
function of the operator associated with the minimal scalar and will
be of the order of
\be
{\rm Im} \langle {\cal O}(k) {\cal O}^\dagger(k) \rangle \sim 
\exp(-\int dU \sqrt{V(U)}).
\ee
If we turn off the non-commutativity effect by setting $\theta =0$
then $V\sim 1/U^2$ and we find that $\int \sqrt{V}$ has logarithmic
divergence at large $U$.  Therefore, the imaginary part is a
polynomial of the ratio between the momentum and the cut-off. This, of
course, is just the WKB approximation to the standard two point
function computation in $AdS_5 \times S_5$ in momentum space
\cite{Gubser:1998bc}.

The case with non-vanishing $|k \theta|$ is the interesting case from
the point of view of the non-commutative gauge invariant operators. As
long as $k^2 <0$, there will be a classically allowed region near
$U=0$, and one can compare the tunneling amplitude to the imaginary
part of the two point function as we did above for $AdS$. There is one
main difference.  The potential does not approach zero at infinity.
Therefore, when we terminate the space at some large $U =
\sqrt{\lambda} \Lambda$ the tunneling amplitude will be exponentially
suppressed.  Keeping the leading terms we get
\be \exp\left( -{1 \over 2 \pi} \sqrt{\lambda} | k \theta | \Lambda +
{2 \pi \over \Gamma({1 \over 4})^2} \sqrt{\sqrt{\lambda} |k \theta|
|k|} \right). \ee

\begin{figure}
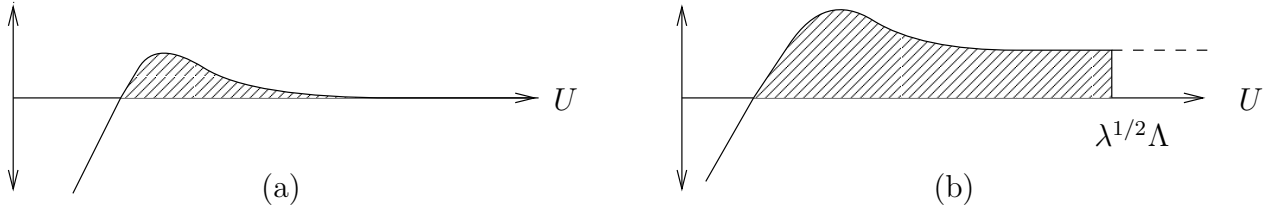

\centerline{\input barrier.pstex_t}
\caption{WKB approximation to the absorption cross section.  (a) In
AdS the potential falls at infinity like $1/U^2$ and hence the
absorption cross section is suppressed like a power of the cutoff.
(b) In the NC version of AdS the potential goes to a constant at
infinity and so the absorption cross section is suppressed like an
exponential of the cutoff.  \label{figd}}
\end{figure}

It is very interesting that this quantity behaves identically to the
field theory result (\ref{universal}) up to replacing $\lambda$ at
weak coupling by $\sim\sqrt{\lambda}$ at strong coupling.  It is
especially interesting that the field theory divergences due to
diagram (a) in figure \ref{figb} have such a simple role on the
supergravity side.  As was mentioned in the previous section,
replacing the 't Hooft constant with its square root is a familiar
effect of strong coupling which we have encountered many times
before. The fact that the supergravity agrees with the field theory
result, up to the 't Hooft coupling, is a strong indication that we
have identified the correct supergravity counterpart of the
correlation function of Wilson lines on the field theory side.

The universality can be seen by turning on excitations on $S^5$ which
correspond to changing the R-charge of the operator.  A short
calculation shows that this will only change the power of momentum in
front of the universal exponent.

\section{Higher point functions}

So far, we have considered only the two point function. Although we
have found a remarkable agreement between the field theory and
supergravity calculations, there is some cause for caution in
interpreting such a result.  In ordinary field theories, the relative
normalization of operators such as $\tr F^2(k)$ for different values
of $k$ is fixed completely by the fact that it is the Fourier
transform of some definite operator in position space.  This, however,
is not the case in non-commutative gauge theories.  As we saw in
section 2, gauge invariant operators in NCSYM contain a piece
corresponding to a Wilson line of definite length which depends on the
momentum $k$. In principle, nothing prevents one from choosing to
normalize these operators in an arbitrarily $k$ dependent
manner. Unlike the commutative theories, operators like $\tr \tilde
F^2(k)$ for different values of $k$ are really different operators. By
changing the normalization of the operators, however, one can conclude
anything about the $k$-dependence of two point function of any
operators
\be \langle \tilde {\cal O}(k) \tilde {\cal O}^\dagger(k) \rangle \rightarrow
f(k)^2 \langle  \tilde {\cal O}(k)\tilde {\cal O}^\dagger(k)
\rangle. \ee

To avoid the ambiguities associated with operator normalizations, one
should consider instead the ratio of $n$-point function to product of
two-point functions of the form
\be\label{wh} \frac{\langle \tilde{\cal O}_1(k_1)\tilde{\cal
O}_2(k_2)\ldots \tilde{\cal O}_n(k_n) \rangle}
{\sqrt{\langle\tilde{\cal O}_1(k_1)\tilde{\cal O}_1^\dagger
(k_1)\rangle \langle\tilde{\cal O}_2(k_2)\tilde{\cal
O}_2^\dagger(k_2)\rangle \ldots \langle\tilde{\cal
O}_n(k_n)\tilde{\cal O}_n^\dagger(k_n)\rangle }}. \ee
It is clear that (\ref{wh}) is invariant with respect to the change in
the normalization of the operators. This is the appropriate quantity
to use in comparing non-commutative and ordinary gauge theories.

In ordinary field theories, (\ref{wh}) will scale like a power for
large $k$.  What happens to (\ref{wh}) for non-commutative theories?
We saw in equation (\ref{universal}) that the two point function of
operators whose normalization are fixed by (\ref{5}) decays
exponentially with respect to the cut-off $\Lambda$ and grows
exponentially with respect to $|k| |\theta k|$. How do the $n$-point
functions of the operators with the same normalization scale? The
$n$-point functions contain an identical dependence on the cut-off
$\Lambda$ due to the UV divergence of gluon propagator starting and
ending on the same Wilson line. Therefore, the dependence on $\Lambda$
will cancel out in (\ref{wh}).

What about the exponential dependence on $|k| |\theta k|$?  This was
due to the resummation of ladder diagrams in the case of the two-point
function. Can something similar happen to the $n$-point function?  The
answer is no.  To see this, note that the origin of the exponential
dependence in the two point function can be traced to the fact that
the Wilson lines are parallel. Since the distance between two parallel
lines is a constant, the contribution from the gluon exchange
propagator gets enhanced by the size of the Wilson line, which gets
larger as we increase the momentum. In the $n$-point function,
however, the lines are not generically parallel, and the contribution
from the propagators are suppressed when the distances between the
points on the Wilson line gets large, even when the Wilson line itself
is very long.

Some sample calculation of the three point function of the Wilson
lines are presented in Appendix B.  Unlike the case of the two point
function, perturbative expansion appears to be controlled in the UV by
the dimensionless parameter $g^2 N / k^2 \theta$ which is certainly a
small parameter. Therefore, there is no need to take the contribution
of the higher order diagrams into consideration. In other words, the
$n$-point function will not grow exponentially, contrary to the
behavior of the two-point functions.

Consequently, the ratio between the $n$-point function and the two
point function is suppressed in the UV like,
\be\label{w}
\frac{\langle \tilde{\cal O}_1(k_1)\tilde{\cal O}_2(k_2)\ldots
\tilde{\cal O}_n(k_n)\rangle}
{\sqrt{\langle\tilde{\cal O}_1(k_1)\tilde{\cal O}^\dagger_1(k_1)\rangle
\langle\tilde{\cal O}_2(k_2)\tilde{\cal O}^\dagger_2(k_2)\rangle
\ldots\langle\tilde{\cal O}_n(k_n)\tilde{\cal O}^\dagger_n(k_n)\rangle}}
\sim \exp \left( - \sum_{i=1}^{n} \sqrt{\frac{g^2 N |k_i \theta |
|k_i|}{16\pi}} \right) .  \ee

What we have found here is an intriguing exponential suppression of
the ratio (\ref{wh}) of $n$-point function to the product of the two
point functions. There are absolutely no ambiguities in the evaluation
of this quantity, and the behavior of this quantity in non-commutative
theory exhibits a clear departure from the behavior of the commutative
theory.

\section{Comparison to string theory}

In the previous section, we described a universal feature of the gauge
invariant observables of non-commutative gauge theories: the ratio of
$n$-point function to the product of two-point functions are
suppressed exponentially at large momentum. This is strongly
reminiscent of the behavior of form factors in high energy fixed angle
scattering of string theory \cite{gm}, and seems like a tantalizing
hint for a connection between string theory and field theories on
non-commutative spaces.

There have been other hints of relation between string theory and
non-commutative field theories. Both are theories with intrinsic
non-locality scale set by a dimensionful parameter: $\alpha'$ in the
case of string theory and $\theta$ in the case of non-commutative
gauge theories. In fact, these two scales are intimately related since
under the open string/closed string map of \cite{sw}, objects whose
size are of the order of $\alpha'$ in closed string metric has the
size of the order of $\theta$ in the open string metric.
Consequently, the string scale in the bulk corresponds to the
non-commutativity scale at the boundary.

In the analysis of the $n$-point function of gauge invariant
observables in non-commutative gauge theories, one finds further
tantalizing similarities to string theory:

\nl $\bullet$ Both string theory and non-commutative gauge theory
exhibits universality in the exponential suppression at large
momentum. On the string theory side, the exponential suppression of
the scattering amplitude is independent of the choice of the vertex
operator. On the non-commutative gauge theory side, the exponential
suppression of (\ref{w}) is independent of the choice of operators
${\cal O}_i$.

\nl $\bullet$ Exponential suppression at large $k$ can be thought of
as an on-shell phenomenon for both string theory and non-commutative
gauge theory. On the string theory side, this is obvious since one is
referring to the behavior of the $S$-matrix. On the gauge theory side,
this may seem strange since (\ref{w}) is an off-shell correlation
function. However, in light of AdS/CFT correspondence which maps
off-shell quantities on the boundary to on-shell quantities on the
bulk, one can think of (\ref{w}) as an on-shell observable of the bulk
theory.

\nl $\bullet$ In both cases fixing the angle while taking the UV limit
is essential to get the exponential suppression.  In string theory
this is well known. On the field theory side, this corresponds to
fixing the relative orientations of $k_i$ as we take $k_i$ to be
large, which fixes the orientation of the Wilson lines to be
non-parallel.

\nl $\bullet$ Both string theory and non-commutative gauge theory
requires resummation of diagrams in order for the universal large $k$
behavior to manifest itself. On the non-commutative gauge theory side,
the resummation of ladder diagrams was necessary in order to obtain
the suppression by an exponential factor $\exp(-|k|)$. On the string
theory side, at each order in the world sheet genus expansion, the
amplitudes are suppressed by $\exp(-k^2)$. The Borel resummation of
these genus expansion amplitudes \cite{om}, valid for $ \alpha' k^2 \gg
\log(1/g)$, gives rise to a suppression by $\exp(-|k|)$ in agreement
with the field theory result.

These similarities are indeed suggestive of a deep connection between
the exponential suppression found in (\ref{w}) and the exponential
suppression of high-energy fixed angle scattering in string theory.
However, there is one crucial difference which immediately rules out
the correspondence, at least in its most naive form.  The point is
that in string theory the suppression is due to large relative
momentum of the different particles involved in the scattering.  This
corresponds to large momentum on the {\em internal} legs.  Here on the
other hand the suppression is due to large momentum on the {\em
external} legs. In this sense, exponential suppression of (\ref{w}) is
quite different from the exponential suppression of high energy fixed
angle scattering of strings.  In the language of supergravity dual,
the exponential suppression of (\ref{w}) is {\it not} due to
Gross-Mende like behavior in the bulk. Instead, it is due to the
propagation from the boundary to the bulk in the geometry given by
(\ref{lk}). Clearly, it would be very interesting to better understand
the physical meaning of the ``suppression by external legs.''

\section{Discussion}

The aim of the present paper is to initiate a systematic treatment of
gauge invariant operators in non-commutative gauge theories.  We saw
that there is a simple and natural generalization of gauge invariant
operators that carry momentum. The loss of gauge invariance due to the
non-zero momentum was compensated by attaching a Wilson line of
definite size. Simple straight Wilson line does the job. This is a
very natural generalization since the operators reduce to ordinary
gauge invariant local operators in the IR. Though formally this is a
very simple modification, we found that it has drastic consequences in
the UV. For example the Wilson line gives rise to a universal
enhancement factor to the two-point function which depends
exponentially on $|k|$, $|k\theta|$ and the cutoff $\Lambda$.  Exactly
the same dependence on these parameters was found from the dual
supergravity description.

The $n$-point functions (with $n>2$) do not receive such an
exponential enhancement and hence the ratio between the $n$-point
function and the two-point function is exponentially suppressed in the
UV.  This looks very similar at first sight to the exponential
dependence of high energy fixed angle string scattering
amplitudes. Closer look revealed, on the contrary, that this is a
completely different effect. The suppression is due to the external
leg factors rather than the internal leg factors.

Clearly, many questions remain to be answered.  For example, to make
further progress it seems important to understand better the mixing
issue both from the field theory side (due to the Wilson lines) and
from the supergravity side (due to the non-trivial background fields).
Understanding the mixing issue is essential to extract the precise
mapping of the degrees of freedom.  It might also be important for the
comparison to string theory.  A somewhat related question is the
interpretation of eq.~(\ref{w}) from the point of view of the
supergravity dual. Since (\ref{w}) does not depend on the
normalization, it should also have an unambiguous formulation in the
language of supergravity.  This can be useful for understanding better
the way holography works in that background which is somewhat similar
to flat space-time.  Refs.~\cite{dani,dasghosh} seems to provide a
good starting point to study that question.

A different kind of question is the relation between the open Wilson
lines considered here and the supergravity description of the Wilson
lines (in the sense of string world sheet minimal surfaces
\cite{rey,malda,dgo}).  In \cite{yaron} it was shown that to reach the
boundary the fundamental strings must carry momentum.  Even though
these Wilson lines involve the scalar fields (unlike the ones
considered in this article), they transform in a similar manner with
respect to the gauge group, and it is natural to expect a close
relation between the two.

It should also be very interesting to try to generalize the gauge
invariant operators of non-commutative gauge theories to string field
theory.  Perhaps the generalization is simpler in the large $B$ field
limit where the string field theory algebra factors out
\cite{Dasgupta:2000ft,chicago,witten2000}.  A different approach to
learn about gauge invariant operators in string theories might be to
consider the S-dual description of the operators considered here which
should be the natural operators in NCOS theories \cite{nc,nc2,nc3}.

\section*{Acknowledgements}

We would like to thank J.~de~Boer, N.~Drukker, I.~Klebanov and
J.~Maldacena for discussions. We also thank the Aspen Center for
Physics where part of that work was done.  The work of D.J.G and A.H
is supported by the NSF under grant No.  PHY94-07194.  The work of N.I
is supported by the NSF under grant No. PHY97-22022.

\section*{Appendix A: Ladder diagrams for the Wilson line two point function}
         {\setcounter{section}{1} \gdef\thesection{\Alph{section}}}
          {\setcounter{equation}{0}}

In this appendix, we will explain how one systematically computes
the ladder diagrams contributing to the $\langle W(k) W^\dagger(k) \rangle$
correlation function, where
\be W(k) = \int d^4 x \, P_* e^{i g \int_0^L d\z A(x+\z)} * e^{i k x}\ . \ee
We begin by expanding these operators
\be  W(k)  = \sum_n (i g)^n \int d^4 x\, \int_{\z_n > \z_{n-1} >  \ldots \z_1}
d\z_i  \, A(x+\z_1) * A(x+\z_2) * \ldots * A(x+\z_n) * e^{i k x} \ee
  \be W^\dagger(k)  = \sum_n (-i g)^n \int d^4 x\, \int_{\z'_n > 
\z'_{n-1} >  \ldots
\z'_1}  d\z'_i  \, A(x+\z'_n) * A(x+\z'_{n-1}) * \ldots * A(x+\z'_1) * 
e^{-i k x} \ .
\ee 
This can be rewritten in momentum space as
\begin{eqnarray} W(k)& = &\sum_n (ig)^n \int d^4 x \int {d^4 p_i 
\over (2 \pi)^4}
\int d\z_i  \\
&&\qquad  A(p_1) e^{-i p_1 (x+\z_1)} * A(p_2) e^{-i p_2 (x + \z_2)} * \ldots
*A(p_n) e^{-i p_n ( x+\z_n)} * e^{i k x} \nonumber \\
W^\dagger(k) &=& \sum_n (-ig)^n \int d^4 x \int {d^4 p_i \over (2 \pi)^4} \int
d\z'_i  \\
&& \qquad A(-p_n) e^{i p_n (x+\z'_n)} * A(-p_{n-1}) e^{i p_{n-1} (x + 
\z'_{n-1})} *
\ldots * A(-p_1) e^{i p_1 ( x+\z'_1)} * e^{-i k x} \ .\nonumber
\end{eqnarray}
Taking the $*$ product and integrating over $x$, we find
\begin{eqnarray}
  W(k) &= &\sum_n (ig)^n \int {d^4 p_i \over (2 \pi)^4} \int d\z_i (2 \pi)^4
\delta(\sum p_i - k) \nonumber \\
&& \qquad  A(p_1) e^{-i p_1 \z_1} A(p_2) e^{-i p_2 \z_2} \ldots 
A(p_n) e^{-i p_n
\z_n} e^{i \sum_{i<j} p_i \theta p_j / 2} \\
W^\dagger(k) &=& \sum_n (-ig)^n \int {d^4 p_i \over (2 \pi)^4} \int 
d\z'_i (2 \pi)^4
\delta(\sum p_i -k)  \nonumber \\
&&\qquad  A(-p_n) e^{i p_n \z'_n} A(-p_{n-1}) e^{i p_{n-1} \z'_{n-1}} 
\ldots A(-p_1)
e^{i p_1 \z'_1} e^{-i \sum_{i<j} p_i \theta p_j / 2} \ .
\end{eqnarray}
Contracting the $A$'s that form the ladder diagram, we find
\be \langle W(k) W^\dagger(k) \rangle  =   g^{2n} N^n \int d\z_i \int 
{d^4 p_i \over (2
\pi)^4}  (2 \pi)^4 \delta(\sum p_i - k) {1 \over p_1^2} {1 \over 
p_2^2 } \ldots {1
\over p_n^2} e^{-i \sum_i p_i (\z_i - \z'_i)}  \ . \ee
Note that the factors of $\exp(i p_i \theta p_j) $ canceled out as to
be expected for a planer diagram.

\begin{figure}
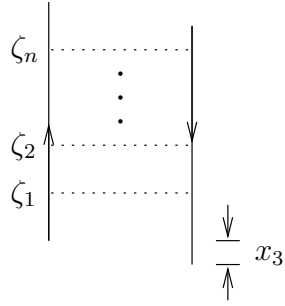

\centerline{\input ladder.pstex_t}
\caption{
The ladder diagrams dominate at large momentum due to the
integration over $\z_i$ which grows linearly with the size of the
line.\label{fige}}
\end{figure}

We can now re-express the correlation function in position  space
\be \langle W(k) W^\dagger(k) \rangle =  g^{2n} N^n \int d^4 x  d 
\z_i\, \prod_i \left(
1 \over 4 \pi^2 (r^2+(x_3+\z_i - \z'_i)^2 )\right) e^{i k r}
\ee
where
\be r^2 = x_0^2 + x_1^2+x_2^2 \ . \ee
Now let us do the $\z_i$, $\z'_i$, and the $x_3$ integrals. We will
first do the $\z'$ integral by simply noting the fact that they
receive most of their contribution from the region
\be x_3+\z_i - \z'_i = 0 \ee
or
\be \z'_i = x_3 + \z_i \ . \ee
Since $0 < \z'_i < L$, we find that $x_3$ must satisfy the condition
\be -\z_1 < x_3 < L-\z_n \ee
in order for all the $\z'_i$ integrals to contribute (See figure
\ref{fige}). Each $\z'_i$ integral will contribute a factor
\be  \int d \z'_i {1 \over 4 \pi^2(r^2 + {\z'_i}^2)} = {1 \over 4 \pi |r|} \ee
where we have used the fact that, for $L \gg k^{-1}$, the integral over
$\z$ can be approximated by an integral over the entire real axis.
The rest of the $\z_i$ and $x_3$ integrals are relatively straight
forward
\be\int_0^L d \z_n \int_0^{\z_n} d \z_{n-1} \ldots \int_0^{\z_2} d \z_1
\int_{-\z_1}^{L-\z_n} dx_3 = {2 L^{n+1} \over (n+1)!} \ .
\ee
This then shows that the correlation function can be expressed in the form
\be \langle W(k) W^\dagger(k) \rangle =  g^{2n} N^n \int d^3 r 
\left({1\over 4 \pi |r|
} \right)^n {2 L^{n+1} \over (n+1)!} e^{i k r} \ .
\ee

\section*{Appendix B: Some three point functions}
         {\setcounter{section}{2} \gdef\thesection{\Alph{section}}}
	 {\setcounter{equation}{0}}

\begin{figure}
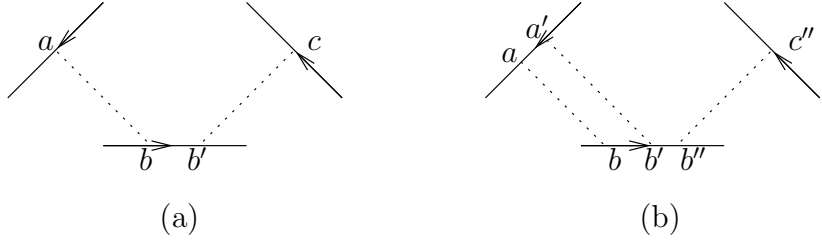

\centerline{\input 3pt.pstex_t}
\caption{A typical three point diagrams. Since the lines are not parallel
(b) will not  dominate over (a)  even at large momenta.\label{figf}}
\end{figure}

In this appendix, we will describe the computation of some diagrams
that contribute to the three point function of the Wilson line
operators in non-commutative gauge theory. We will consider the
diagrams illustrated in figure \ref{figf}.
We will begin by considering the diagram (a) in figure \ref{figf}.
The terms in the perturbative expansion of the Wilson line operators
which contribute to this diagram are

\begin{eqnarray}
W_1 &=& \int d^4 x  \, da \, \z_1' \inn  A (x+\z_1(a)) * e^{i k_1 x} 
= \int da\,   \z_1' \inn  A(k) e^{-i k_1 \z_1(a)} = \z_1' \inn 
A(k_1)  \\
W_2 & = & \int d^4 x \, db \, db'\ \z_2' \inn  A(x+\z_2(b)) * \z_2' 
\inn  A (x+\z_2(b')) * e^{i k_2 x} \\
     & = & \int db\, db'\,  \int  {d^4 p \over (2 \pi)^4} {d^4 q \over 
(2 \pi)^4} \z_2' \inn  A (p) \, \z_2' \inn  A(q) e^{i p \theta q/2 - 
i p \z_2(b) - i q \z_2(b') } (2 \pi)^4 \delta^4(p+q - k_2) \nonumber 
\\
W_3 &  = & \int dc \ \z_3' \inn  A(k_3) e^{-i k_3 \z_3(c)} = \z_3' \inn  A(k_3)
\end{eqnarray}
so that
\be \langle W_1 W_2 W_3 \rangle  = \int db \, db'\,{\z_1' \inn \z_2' 
\z_3' \inn \z_2' \over k_1^2 k_3^2} e^{i k_1 \theta k_3 /2 + i k_1 
\z_2(b) + i k_3 z_2(b') } \,
\ee
where $a$, $b$, $b'$ and $c$ take values between 0 and 1, and
\be \z_1(a) = k_1 \theta a \qquad \z_2(b) = k_2 \theta b  \qquad \z_3(c) = k_3
\theta c \ .\ee
Therefore, we have
\be \langle W_1 W_2 W_3 \rangle  = \int db \, db'\,{(k_1 \theta^2 
k_2)(k_3 \theta^2 k_2) \over k_1^2 k_3^2} e^{i k_1 \theta k_3 /2 + i 
k_1 \theta k_2 b + i k_3 \theta k_2 b' } \ .
\ee
Using
\be k_1 \theta k_2 = - k_1 \theta k_3, \qquad k_3 \theta k_2 =  - k_2 
\theta k_3 = k_1 \theta k_3 \ee
this becomes
\be \langle W_1 W_2 W_3 \rangle  = \int db \, db'\,{(k_1 \theta^2 
k_2)(k_3 \theta^2 k_2)  \over k_1^2 k_3^2} e^{i k_1 \theta k_3 /2 - i 
k_1 \theta k_3 (b-b')} \ .
\ee
Doing the  integral over $b$ and $b'$, we find
\be \langle W_1 W_2 W_3 \rangle  = {(k_1 \theta^2 k_2)(k_3 \theta^2 
k_2)  \over k_1^2 k_3^2} e^{i k_1 \theta k_3 /2 } \left( {i \over k_1 
\theta k_3} + {1 - e^{i k_1 \theta k_3} \over (k_1 \theta k_3)^2} 
\right) \ .
\ee
For large $k^2 \theta$, the first term in the parenthesis dominates,
and its overall scaling with respect to $k$ and $\theta$ is
\be \langle W_1 W_2 W_3 \rangle \approx g^4 N^2 {\theta^3 \over k^2}  \ .
\label{3pta} \ee

Let us now consider the contribution from diagram (b) in figure
\ref{figf}.  The operators are
\begin{eqnarray}
W_1 & = & \int d^4 x \int da \int^a da' \z_1' \inn A(x+\z_1(a'))* 
\z_2' \inn A(x+\z(a)) * e^{i k_1 x} \\
& = &  \int da \int da' \int {d^4 p \over (2 \pi)^4} \int {d^4 q 
\over (2 \pi)^4} \z_1' \inn A(q) \z_1' \inn A(p) e^{i q \theta p /2 - 
i q \z_1(a') - i p \z_1(a)} (2\pi)^4 \delta^4(p+q-k_1) \nonumber  \\
W_2 & = & \int d^4 x \int db'' \int^{b''} d b' \int^{b'} db \nonumber \\
&& \qquad \z_2' \inn A(x+\z_2(b)) * \z_2' \inn A(x+\z_2(b')) * \z_2' 
\inn A(x+\z_2(b'')) * e^{i k_2 x} \nonumber \\
& = & \int db'' \int db' \int db \int{d^4 p \over (2 \pi)^4} \int{d^4 
q \over (2 \pi)^4} \int{d^4 r \over (2 \pi)^4} (2 \pi)^4 \delta^4 
(k_2+p+q+r) \\
&&  \z_2' \inn A(-p)\, \z_2' \inn  A(-q)\, \z_2' \inn A(-r) e^{i p 
\theta q/2 + i p \theta r /2 + i q \theta r/2 + i p \z_2(b) + i q 
\z_2(b') + i r \z_2(b'')}  \nonumber \\
W_3 & = & \z_3' \inn  A(k_3) \ .
\end{eqnarray}

Then $\langle W_1 W_2 W_3 \rangle$ goes as
\be\int da da' db'' db' db \int {d^4 q \over (2 \pi)^4} {1 \over q^2} 
{1 \over (k_1-q)^2} e^{-i q a' - i (k_1-q) a + i (k-q) b + i q b' - i 
k_3 b''} e^{i k_1 \theta k_3/2} {(\z_1' \inn \z_2')^2 \z_2' \inn 
\z_3' \over k_3^2} \ .
\ee
This can be written in the form
\be  \langle W_1 W_2 W_3 \rangle = \int {d^4 q \over (2 \pi)^4} {1 
\over q^2} {1 \over (k_1-q)^2} F(q) e^{i k_1 \theta k_3/2} {(\z_1' 
\inn \z_2')^2 \z_2' \inn \z_3'  \over k_3^2} \ ,
\ee
where
\be F(q) = \int da da' db'' db' db\,  e^{-i q \theta k_1 a' - i 
(k_1-q) \theta k_1 a + i (k_1-q) \theta k_2 b + i q \theta k_2 b' - i 
k_3 \theta k_2 b''} \ . \ee
Since $F(q)$ is again damping function with a width of the order $q
\approx 1/ k \theta$, to leading order we can approximate
\be  \langle W_1 W_2 W_3 \rangle = \int  {1 \over k_1^2} \left(\int 
d^2 q F(q) \right) e^{i k_1 \theta k_3/2} {(\z_1' \inn \z_2')^2 \z_2' 
\inn \z_3'  \over k_3^2} \ . \label{aaaa}
\ee
The $d^2 q$ integral will give rise to a $\delta$-function. One finds
that
\be \int d^2 q F(q) = \int da da' db'' db' db \ \delta^2( \theta k_1
(a'-a) + \theta k_2 (b-b')) e^{i k_1 \theta k_2 b - i k_3 \theta k_2
b''} \ . \ee
This will go like
\be \int d^2 q F(q) \approx {1 \over (\theta k)^2} {1 \over (k \theta
k)^2} \approx {1 \over k^6 \theta^4} , \ee
where the first factor comes from integrating over the $\delta$
function and the second factor comes from integrating over the phase
factors. Plugging this back into (\ref{aaaa}) leads to the conclusion
that
\be \langle W_1 W_2 W_3 \rangle \approx g^6 N^3 {\theta^2 \over k^4}  \ .
\label{3ptb}\ee
Comparing (\ref{3pta}) and (\ref{3ptb}), we find that the
dimensionless parameter controlling the perturbative expansion is $g^2
N / k^2 \theta$.

\begingroup\raggedright\endgroup
\end{document}

%% file: operators.pstex_t
\begin{picture}(0,0)%
\epsfig{file=operators.pstex}%
\end{picture}%
\setlength{\unitlength}{3552sp}%
\begingroup\makeatletter\ifx\SetFigFont\undefined%
\gdef\SetFigFont#1#2#3#4#5{%
  \reset@font\fontsize{#1}{#2pt}%
  \fontfamily{#3}\fontseries{#4}\fontshape{#5}%
  \selectfont}%
\fi\endgroup%
\begin{picture}(8552,2327)(150,-3437)
\put(6383,-3361){\makebox(0,0)[lb]{\smash{\SetFigFont{11}{13.2}{\rmdefault}{\mddefault}{\updefault}(b)}}}
\put(1853,-3361){\makebox(0,0)[lb]{\smash{\SetFigFont{11}{13.2}{\rmdefault}{\mddefault}{\updefault}(a)}}}
\put(5401,-2055){\makebox(0,0)[lb]{\smash{\SetFigFont{11}{13.2}{\rmdefault}{\mddefault}{\updefault}$\ne$}}}
\put(346,-2085){\makebox(0,0)[lb]{\smash{\SetFigFont{11}{13.2}{\rmdefault}{\mddefault}{\updefault}$k \theta$}}}
\put(7276,-2056){\makebox(0,0)[lb]{\smash{\SetFigFont{11}{13.2}{\rmdefault}{\mddefault}{\updefault}=}}}
\put(1898,-2056){\makebox(0,0)[lb]{\smash{\SetFigFont{11}{13.2}{\rmdefault}{\mddefault}{\updefault}=}}}
\end{picture}

%% file: feynman.pstex_t
\begin{picture}(0,0)%
\epsfig{file=feynman.pstex}%
\end{picture}%
\setlength{\unitlength}{0.00083300in}%
\begingroup\makeatletter\ifx\SetFigFont\undefined%
\gdef\SetFigFont#1#2#3#4#5{%
  \reset@font\fontsize{#1}{#2pt}%
  \fontfamily{#3}\fontseries{#4}\fontshape{#5}%
  \selectfont}%
\fi\endgroup%
\begin{picture}(6602,1888)(1050,-3137)
\put(1535,-3113){\makebox(0,0)[lb]{\smash{\SetFigFont{12}{14.4}{\rmdefault}{\mddefault}{\updefault}(a)}}}
\put(3321,-3113){\makebox(0,0)[lb]{\smash{\SetFigFont{12}{14.4}{\rmdefault}{\mddefault}{\updefault}(b)}}}
\put(5108,-3113){\makebox(0,0)[lb]{\smash{\SetFigFont{12}{14.4}{\rmdefault}{\mddefault}{\updefault}(c)}}}
\put(6895,-3113){\makebox(0,0)[lb]{\smash{\SetFigFont{12}{14.4}{\rmdefault}{\mddefault}{\updefault}(d)}}}
\end{picture}

%% file: loop.pstex_t
\begin{picture}(0,0)%
\epsfig{file=loop.pstex}%
\end{picture}%
\setlength{\unitlength}{0.00083300in}%
\begingroup\makeatletter\ifx\SetFigFont\undefined%
\gdef\SetFigFont#1#2#3#4#5{%
  \reset@font\fontsize{#1}{#2pt}%
  \fontfamily{#3}\fontseries{#4}\fontshape{#5}%
  \selectfont}%
\fi\endgroup%
\begin{picture}(3913,2402)(1939,-2537)
\put(2048,-1612){\makebox(0,0)[lb]{\smash{\SetFigFont{12}{14.4}{\rmdefault}{\mddefault}{\updefault}$T$}}}
\put(2810,-428){\makebox(0,0)[lb]{\smash{\SetFigFont{12}{14.4}{\rmdefault}{\mddefault}{\updefault}$\!L$}}}
\put(4351,-1618){\makebox(0,0)[lb]{\smash{\SetFigFont{12}{14.4}{\rmdefault}{\mddefault}{\updefault}$k \theta$}}}
\put(5130,-413){\makebox(0,0)[lb]{\smash{\SetFigFont{12}{14.4}{\rmdefault}{\mddefault}{\updefault}$\!1/k$}}}
\end{picture}

%% file: barrier.pstex_t
\begin{picture}(0,0)%
\epsfig{file=barrier.pstex}%
\end{picture}%
\setlength{\unitlength}{0.00083300in}%
\begingroup\makeatletter\ifx\SetFigFont\undefined%
\gdef\SetFigFont#1#2#3#4#5{%
  \reset@font\fontsize{#1}{#2pt}%
  \fontfamily{#3}\fontseries{#4}\fontshape{#5}%
  \selectfont}%
\fi\endgroup%
\begin{picture}(7813,1288)(289,-2537)
\put(7951,-1936){\makebox(0,0)[lb]{\smash{\SetFigFont{12}{14.4}{\rmdefault}{\mddefault}{\updefault}$U$}}}
\put(3676,-1936){\makebox(0,0)[lb]{\smash{\SetFigFont{12}{14.4}{\rmdefault}{\mddefault}{\updefault}$U$}}}
\put(1853,-2476){\makebox(0,0)[lb]{\smash{\SetFigFont{12}{14.4}{\rmdefault}{\mddefault}{\updefault}(a)}}}
\put(6046,-2476){\makebox(0,0)[lb]{\smash{\SetFigFont{12}{14.4}{\rmdefault}{\mddefault}{\updefault}(b)}}}
\put(7051,-2161){\makebox(0,0)[lb]{\smash{\SetFigFont{12}{14.4}{\rmdefault}{\mddefault}{\updefault}$\lambda^{1/2} \Lambda$}}}
\end{picture}

%% file: ladder.pstex_t
\begin{picture}(0,0)%
\epsfig{file=ladder.pstex}%
\end{picture}%
\setlength{\unitlength}{0.00083300in}%
\begingroup\makeatletter\ifx\SetFigFont\undefined%
\gdef\SetFigFont#1#2#3#4#5{%
  \reset@font\fontsize{#1}{#2pt}%
  \fontfamily{#3}\fontseries{#4}\fontshape{#5}%
  \selectfont}%
\fi\endgroup%
\begin{picture}(1802,1899)(6300,-3148)
\put(6376,-1606){\makebox(0,0)[lb]{\smash{\SetFigFont{12}{14.4}{\rmdefault}{\mddefault}{\updefault}$\zeta_n$}}}
\put(6376,-2206){\makebox(0,0)[lb]{\smash{\SetFigFont{12}{14.4}{\rmdefault}{\mddefault}{\updefault}$\zeta_2$}}}
\put(6376,-2506){\makebox(0,0)[lb]{\smash{\SetFigFont{12}{14.4}{\rmdefault}{\mddefault}{\updefault}$\zeta_1$}}}
\put(7899,-2871){\makebox(0,0)[lb]{\smash{\SetFigFont{12}{14.4}{\rmdefault}{\mddefault}{\updefault}$x_3$}}}
\end{picture}

%% file: 3pt.pstex_t
\begin{picture}(0,0)%
\epsfig{file=3pt.pstex}%
\end{picture}%
\setlength{\unitlength}{0.00083300in}%
\begingroup\makeatletter\ifx\SetFigFont\undefined%
\gdef\SetFigFont#1#2#3#4#5{%
  \reset@font\fontsize{#1}{#2pt}%
  \fontfamily{#3}\fontseries{#4}\fontshape{#5}%
  \selectfont}%
\fi\endgroup%
\begin{picture}(5124,1438)(1489,-5087)
\put(1726,-3961){\makebox(0,0)[lb]{\smash{\SetFigFont{12}{14.4}{\rmdefault}{\mddefault}{\updefault}$\!a$}}}
\put(2326,-4711){\makebox(0,0)[lb]{\smash{\SetFigFont{12}{14.4}{\rmdefault}{\mddefault}{\updefault}$b$}}}
\put(2626,-4711){\makebox(0,0)[lb]{\smash{\SetFigFont{12}{14.4}{\rmdefault}{\mddefault}{\updefault}$b'$}}}
\put(3376,-3961){\makebox(0,0)[lb]{\smash{\SetFigFont{12}{14.4}{\rmdefault}{\mddefault}{\updefault}$c$}}}
\put(4801,-3886){\makebox(0,0)[lb]{\smash{\SetFigFont{12}{14.4}{\rmdefault}{\mddefault}{\updefault}$\!\!a'$}}}
\put(5476,-4711){\makebox(0,0)[lb]{\smash{\SetFigFont{12}{14.4}{\rmdefault}{\mddefault}{\updefault}$b'$}}}
\put(5251,-4711){\makebox(0,0)[lb]{\smash{\SetFigFont{12}{14.4}{\rmdefault}{\mddefault}{\updefault}$b$}}}
\put(5701,-4711){\makebox(0,0)[lb]{\smash{\SetFigFont{12}{14.4}{\rmdefault}{\mddefault}{\updefault}$b''$}}}
\put(6376,-3961){\makebox(0,0)[lb]{\smash{\SetFigFont{12}{14.4}{\rmdefault}{\mddefault}{\updefault}$c''$}}}
\put(4651,-4036){\makebox(0,0)[lb]{\smash{\SetFigFont{12}{14.4}{\rmdefault}{\mddefault}{\updefault}$\!\!a$}}}
\put(2453,-5063){\makebox(0,0)[lb]{\smash{\SetFigFont{12}{14.4}{\rmdefault}{\mddefault}{\updefault}(a)}}}
\put(5446,-5063){\makebox(0,0)[lb]{\smash{\SetFigFont{12}{14.4}{\rmdefault}{\mddefault}{\updefault}(b)}}}
\end{picture}